\documentclass[epj]{svjour}
\usepackage{graphicx}
\usepackage{times}

\def\xe{{\mbox{$x^{\prime}$}}}
\def\ye{{\mbox{$y^{\prime}$}}}
\def\ze{{\mbox{$z^{\prime}$}}}

\def\averad{{\mbox{$\langle \Delta R \rangle$}}}
\def\avevel{{\mbox{$\langle v_{\rm DEP} \rangle$}}}

\def\radlat{{\mbox{$\Delta R_{\rm lat}$}}}

\def\el{{\mbox{$\vec{E}$}}}
\def\dip{{\mbox{$\vec{p}$}}}
\def\ap{{\mbox{$\alpha_p$}}}

\def\fdep{{\mbox{$\vec{F}_{\rm DEP}$}}}
\def\fdepsc{{\mbox{$F_{\rm DEP}$}}}

\def\tdep{{\mbox{$\Theta_{\rm DEP}$}}}
\def\mobdep{{\mbox{$\mu_{\rm DEP}$}}}
\def\nablae2{{\mbox{$\arrowvert \nabla E^2 \arrowvert$}}}

\def\eunit{{\mbox{$\varepsilon/e\sigma$}}}
\def\tunit{{\mbox{$\varepsilon/k_{\rm B}$}}}

\begin{document}

\tolerance=10000 \hyphenpenalty=2000
\voffset=0.4in

\bibliographystyle{prstyx}

\title{
Dielectrophoresis of nanocolloids: a molecular dynamics study
}

\author{E. Salonen\inst{1}\fnmsep\thanks{email: emppu.salonen@hut.fi} \and 
E. Terama\inst{1}\fnmsep\thanks{email: terama@pcu.helsinki.fi} \and
I. Vattulainen\inst{1}\fnmsep\inst{2}\fnmsep\thanks{email: ilpo.vattulainen@hut.fi} \and
M. Karttunen\inst{3}\fnmsep\thanks{email: mikko.karttunen@hut.fi}}

\institute{ 
Biological Physics and Soft Matter Group, Laboratory of Physics and Helsinki
Institute of Physics, Helsinki University of Technology, P.O. Box 1100,
FI-02015 HUT, Finland \and
Memphys-Center of Biomembrane Physics, Physics Department, University
of Southern Denmark, DK-5230 Odense M, Denmark \and
Biophysics and Statistical Mechanics Group, Laboratory of Computational Engineering,
Helsinki University of Technology, P.O. Box 9203, FI-02015 HUT, Finland 
}

\date{Received: November 9, 2004 ~/ Revised version: \today}

\abstract{ 
Dielectrophoresis (DEP), the motion of polarizable particles in
non-uniform electric fields, has become an important tool for the
transport, separation, and characterization of microparticles in
biomedical and nanoelectronics research. In this article we present,
to our knowledge, the first molecular dynamics simulations of DEP of
nanometer-sized colloidal particles. We introduce a simplified model
for polarizable nanoparticles, consisting of a large charged macroion
and oppositely charged microions, in an explicit solvent. The model is
then used to study DEP motion of the particle at different
combinations of temperature and electric field strength. In accord
with linear response theory, the particle drift velocities are shown
to be proportional to the DEP force. Analysis of the colloid DEP
mobility shows a clear time dependence, demonstrating the variation of
friction under non-equilibrium.  The time dependence of the mobility
further results in an apparent weak variation of the DEP displacements
with temperature.
\\
}

\maketitle


\section{Introduction}
\label{intro}

A polarizable particle in a non-uniform electric field is set in
motion due to the coupling between the electric field gradient and the
induced polarization. This effect was named {\it dielectrophoresis}
(DEP) by Pohl in his pioneering studies in the 1950's
\cite{Pohl51,Pohl58}.  DEP can be understood by considering the
simple picture of a dipole in a non-uniform electric field. The dipole
will be at least partly oriented in the direction of the field
gradient and hence one end of the dipole will experience a stronger
electric field strength than the other. This results in a non-zero net
force and the dipole is set in motion.

If the polarizable particle is smaller than the characteristic length
at which the electric field \el{} changes appreciably, the DEP force affecting
the particle can be approximated as a coupling between its induced dipole
moment \dip{} and the field,
\begin{equation}
\fdep = ({\mathbf \dip} \cdot \nabla) \el. 
\label{genforce}
\end{equation}
For an ideal dielectric the induced dipole moment is linearly
proportional to the electric field, \dip{} = \ap\el{}, where \ap{} is
the total effective polarizability of the particle that depends on the
dielectric properties of the particle and the suspending
medium. Combining this relation with Eq. (\ref{genforce}), one obtains
a general equation for the DEP force,
\begin{equation}
\fdep = \frac{1}{2} \ap \nabla \el^2.
\label{depforce}
\end{equation}

Two important observations can be made from
Eq. (\ref{depforce}). First, the direction of the DEP force is
parallel to the gradient of the electric field squared, but does not
depend on the actual polarity of the field. Hence, the same DEP effect
can be achieved with both DC and AC electric fields.  Second, the
direction of the force also depends on the polarizability of the
particle. If the particle is more polarizable than the medium, it will
be attracted to regions of higher field strength. If, on the other
hand, the particle is less polarizable, it will be repelled from the
high-field regions. The first case is commonly referred to as {\it
positive} DEP and the latter as {\it negative} DEP.

With modern microfabrication techniques based on photo- and
electronlithographies it is possible to routinely manufacture smooth
electrode systems in the sub-micron scale. Such miniaturized
structures can generate high electric field strengths and field
gradients with low applied voltages, thus minimizing the unwanted
effects of electrothermally induced motion of the suspending fluid
\cite{Ramos98}.  Microelectrode structures employing DEP have been
used, {\it e.g.}, in the self-assembly of micrometer-sized silicon
resistors \cite{Lee03} and trapping of conducting nanoparticles
\cite{Bezryadin97,Zheng04}, as well as alignment and assembly of
metallic nanowires \cite{Smith00,Hermanson01} and carbon nanotubes
\cite{Yamamoto98,Chan04}. The precision obtained in these studies and
the fact that many particles can be positioned in the electrode
systems in a parallel fashion show that DEP is a very promising tool
for the assembly of nanoelectronics devices on a larger scale (for
recent reviews on the use of DEP in nanotechnology, see
Refs. \cite{Hughes00,Burke04}).  Since DEP is a non-invasive and
non-destructive technique, it has also been used to manipulate
biologically relevant microparticles in various applications
\cite{Markx94,Washizu94,Becker94,Hughes98a,Morgan99,Chou02,Huang02,Gray04}.

In comparison to standard electrophoresis ({\it i.e.}, particle motion
due to the coupling between an applied electric field and a non-zero
net charge), DEP has the disadvantage that the polarization forces
involved are typically quite weak.  In general, efficient particle
manipulation in microelectrode systems requires balancing DEP against
other forces present, such as viscous, buoyancy, and
electrohydrodynamic forces \cite{Ramos98}. Clever experimental setups
have indeed been devised to trap particles of the order of only a few
nm in diameter \cite{Bezryadin97,Zheng04}.  But while many of the
hindrances can be overcome with appropriate choices for suspending
media and electric field configurations, a fundamental limit on the
precision and efficiency of DEP manipulation is set by the thermal
motion of the particles.  For the case of stable particle trapping
over long periods in time, the DEP force clearly has to be much
stronger than the thermal noise from the particle-solvent
interactions.  To this end, reasonable criteria of the trapping
conditions for isolated particles with idealized geometries can be
formulated by using simplified hydrodynamics
\cite{Washizu94,Hughes98a}. However, in the event of particle
aggregation the situation becomes much more complicated
\cite{Huang03,Huang04a}.

Computational modeling has previously been used to study diverse
electrokinetic phenomena, such as electrophoresis
\cite{Tanaka02,Yeh04}, overcharging and like-charge attraction
\cite{Linse99,Messina00,Patra03}, and structure formation in
electrorheological fluids \cite{Klingenberg89,Tao94}.  It is then
natural to ask whether information on DEP motion of nanoparticles
could be obtained from dynamic computer simulations employing
simplified, but still physically sound models for particles and their
interactions.  The motion of nanoscopic particles and particle
aggregates in electrode systems of realistic dimensions is essentially
a multi-scale problem, with important processes taking place at times
ranging from picoseconds (dynamic processes on the molecular scale) to
microseconds (larger aggregate formation), and all the way to seconds
(particle concentration in specific parts of the device). At present
there exists a variety of methods to tackle multi-scale problems at
different levels of accuracy
\cite{Lyubartsev02a,Softsimu2003,Nanoreview04a,Murtola04}.

As the initial step to computational modeling of DEP, in this article
we present, to our knowledge, the first molecular dynamics (MD)
simulations of dielectrophoresis.  The goal of the present work is to
model the DEP of a single colloidal particle at varying combinations
of temperature and applied electric field strength. We verify that the
model exhibits linear drift velocity to DEP force relation and further
study the particle DEP mobility and the temperature dependence of the
motion.

This article is organized as follows. In Section \ref{method}, we will
outline in detail and validate the nanoparticle DEP model used in the
present work, and further examine the extent of finite-size effects in
our model system. Section \ref{results} then comprises the simulation
results on the single-particle DEP transport.  Section
\ref{conclusion} closes the article with the concluding remarks.


\section{Simulation method}
\label{method}

\subsection{General features of the model}
\label{generalfeat}

We have used classical MD simulations to model the dielectrophoresis
of a single charge-neutral colloidal particle in a non-conducting
solvent. As a generic model for a polarizable particle we used a
system consisting of a large sphere, carrying a charge $q_{\rm mac}$,
on which $N_{\rm mic}$ small particles with a charge $q_{\rm mic}$
were electrostatically bound (see Sec. \ref{elstat}). In the remainder
of the text, we adopt the following terminology: the large sphere will
be called {\it macroion}, and the small bound particles {\it
microions}. The macroion-microion complex will be called {\it
colloidal particle} or simply {\it colloid}. As the solvent
particles are smaller than the colloid in linear dimension only by an
order of magnitude, the colloid can be viewed as a nanoparticle
immersed in a molecular solvent.

For a proper description of colloid-solvent interactions
under non-equilibrium conditions, the suspending medium in which the
colloid was immersed was simulated with explicit neutral particles.
Excluded volume interactions between all particle types were modeled
using a truncated and shifted 12-6 Lennard-Jones potential (also
called the WCA potential) \cite{Weeks71}. In order to take into
account the size of the macroion, an additional hard-core radius $r_0$
was included in the denominator of the repulsive and attractive power
terms \cite{Laird89},
\begin{equation}
\Phi_{\rm WCA} = 4 \varepsilon \left[ \left( \frac{\sigma}{r_{ij} -
r_0} \right)^{12} - \left( \frac{\sigma}{r_{ij} -
r_0}\right)^{6}\right] + \varepsilon,
\end{equation}
where $r_{ij} = \arrowvert \vec{r}_j - \vec{r}_i \arrowvert$ is the
center-to-center distance between the interacting particles $i$ and
$j$ ($\vec{r}_i$ being the position vector of particle $i$).  The
cutoff distance of the potential was $r_0 + 2^{1/6} \sigma$, with $r_0
= 4.5 \sigma$ for interaction pairs involving the macroion and $r_0 =
0$ otherwise.  The characteristic parameters $\sigma$ and
$\varepsilon$ of the potential were chosen as the basic units of
length and energy, respectively. With $m$ further designating the unit
of mass, the unit of time is then defined as $\tau = \sigma
\sqrt{m/\varepsilon}$.  

The simulation cell was cubic with periodic boundary conditions and a
side length $L_0 = 35 \sigma$.  The solvent particles and microions
were assigned a mass 1$m$ and the solvent number density was chosen
as $\rho_s$ = 0.3 $\sigma^{-3}$. For the macroion, the mass was chosen
to correspond to the mass of bulk solvent equal to its volume, giving
a value of 155$m$.  Suitable estimates for effective particle sizes
could be obtained from the distances of closest approach between
different particles at temperature $T$, using the relation $\Phi_{\rm
WCA} = k_{\rm B} T$.  Our choice of $r_0$ then gave effective particle
radii roughly $5 \sigma$ for the macroions and $\sigma/2$ for the
microions and solvent particles. Although particle sizes defined this
way depend on the system temperature, the steepness of the WCA
potential slope ensured that the effective particle radii did not vary
significantly in the temperature range used in this study.

\subsection{Electrostatic interactions}
\label{elstat}

The electrostatic interactions between the ions were calculated
directly from Coulomb's law.  The energy associated with the
interaction of particles $i$ and $j$ was then given by 
\begin{equation}
V_c = \frac{q_i q_j}{4 \pi \epsilon_0 \epsilon_r r_{ij}}, 
\end{equation}
where $q_i$ is the charge of particle $i$, $\epsilon_0$ is the vacuum
permittivity, and $\epsilon_r$ the relative permittivity of the
medium. To ensure that the microions stay bound to the macroion
surface we chose the regime of strong Coulomb coupling by setting $(4
\pi \epsilon_0 \epsilon_r)^{-1}$ = 56 $\varepsilon\sigma/e^2$,
independent of temperature and constant in the entire simulation cell.
We did not employ any type of lattice summation or multipole methods
as this would be problematic due to the non-uniformity of the external
electric field (see Sec. \ref{efield}).

The charge $q_{\rm mac}$ = $-20 e$, where $e$ is the elementary
charge, was assigned to the interaction center of the macroion.  To
prevent an electrophoretic component in the external electric
field-induced motion, overall charge neutrality was imposed on the
macroion-microion system, $q_{\rm mac} + N_{\rm mic} q_{\rm mic} =
0$. The number and charge of the microions were then chosen as $N_{\rm
mic}$ = 10 and $q_{\rm mic}$ = $+2e$, respectively.

At a first glance, the choice of spatially uniform value of
$\epsilon_r$ may seem controversial, as the magnitude and direction of
the DEP force generally result from the mismatch of the complex
permittivities of the manipulated particles and the suspending medium.
A purely practical reason for this simplification is the fact that
taking into account the induced surface charges of polarization (image
charges) on the macroion \cite{Messina02b} is computationally
extremely intensive, making dynamic simulations, such as the ones
presented here, dramatically more time-consuming.  However,
considering the picture of a single nanoparticle in solution, it is
not even meaningful to talk about its permittivity.  With our generic
macroion-microion system, the polarization of the colloid originates
simply from the redistribution of the microions on the macroion
surface.

\subsection{External field}
\label{efield}

For the modeling of DEP, it is preferable to use an electric field
geometry which is simple to realize and has a large field gradient in
order to produce appreciable dielectrophoretic motion. In this work we
have used a spherically symmetric DC electric field,
\begin{equation}
\el(R) = E_0 \left( \frac{R_0}{R} \right)^2 \vec{e}_{\mathrm R},
\label{sphericalfield}
\end{equation}
where $E_0$ is the electric field strength at the initial radial
position $R_0$ of the colloid, and $\vec{e}_{\mathrm R}$ is the radial
unit vector. As the colloid moves in the field, $R_0$ further
determines the characteristic length over which the field strength
changes appreciably.  This type of an electric field geometry is a
good approximation for many experimental DEP setups \cite{Burke04},
and was thus a reasonable choice for a model system. The DEP force
arising from the non-uniformity of the field is
\begin{equation}
\fdep = -2 \ap \frac{E_0^2}{R_0} \left( \frac{R_0}{R} \right) ^5
\vec{e}_{\mathrm R},
\label{sphericalforce}
\end{equation}
as results from Eq. (\ref{depforce}). Since our model colloid is
always more polarizable than the surrounding medium, the simulations
corresponded to the case of positive DEP.  For all the simulations
presented in this article we used $R_0 = 1500 \sigma$.

The electric field strength affecting a charged particle was
calculated by using an additional coordinate system, with the origin
set at the force center of the electric field. The vector \vec{R} was
used to denote the position of the center of the microscopic
simulation cell in the electric field coordinates.  The position of a
charged particle $i$ was then \vec{r}$^{\prime}_i$ = \vec{R} +
\vec{r}$_i$. (In the following, all coordinates marked with prime will
refer to the electric field frame of reference.)  We emphasize that
the only effect of assigning the additional electric field coordinates
for the charged particles was to take into account the spatial
dependence of the electric field and its gradient.  During the course
of the simulations the vector \vec{R} was updated, and hence all
\vec{r}$^{\prime}_i$ as well, according to the colloid displacements
in the simulation cell (see below in Sec. \ref{simulations}).

In the absence of an external electric field the microions are evenly
distributed on the surface of the macroion due to their mutual
electrostatic repulsions. When an external field is applied,
polarization of the system is obtained with the redistribution of the
microions.  In order to calculate the DEP force affecting the
colloidal particle, its polarizability \ap{} needed to be
determined. This was done by calculating the induced dipole moment of
the particle, \dip{} = $\sum_i q_i \vec{r}_i$, where the summation is
carried out over the ions, in the following way. The colloid was put
under a selected electric field strength and the microions were
allowed to freely redistribute themselves on the macroion surface.
The total kinetic energy of the system was very slowly quenched to
zero, resulting in a minimum energy state of the polarized
colloid. Note that this type of MD energy minimization cannot
rigorously guarantee that the obtained ion configuration is the true
global energy minimum state of the system \cite{Patra03}. However,
several independent energy minimization simulations showed only
negligible differences in the total energy and thus the obtained
values were assumed to lie very close to the true energy minimum.

As expected, the dipole moment of the macroion-microion system was
observed to be linearly proportional to the electric field strength in
a wide range of $E_0$.  Deviations from the linearity occured at very
low field strengths due to microion-microion electrostatic
correlations and at extremely high field strengths (above 17 \eunit,
which is nearly half the electric field strength due to the macroion
bare charge at a distance of $r_0 + \sigma$), as the microions were
stripped from the macroion surface.  A linear fit to the data gave the
colloid a total polarizability $\ap$ = 4.94 $\pm$ 0.16
$e^2\sigma^2/\varepsilon$. Comparing the DEP forces calculated with
fit value of \ap{} and Eq.  (\ref{depforce}) to the ones obtained
directly from the simulations showed excellent agreement, thus
validating the dipole approximation.

\subsection{DEP simulations}
\label{simulations}

The DEP simulations were carried out as follows. The polarized
colloid, in its minimum energy state, was set at the center of the
simulation cell and the remaining free volume was filled with randomly
distributed solvent particles. In order to avoid strong repulsion
forces due to excluded volume overlaps, it was ensured that all
interparticle distances were larger than the WCA potential cutoff
distance $r_0 + 2^{1/6} \sigma$. The system was then thermalized for
85$\tau$ at a selected temperature ($T = 0.03 - 2.4$ \tunit) by
coupling all the particles to the thermostat by Berendsen {\it et al.}
\cite{Berendsen84} with a coupling time constant of 0.14$\tau$.

The range of temperatures used in this study may sound broad in view
of experiments. However, just like the electric field strength in our
model is a measure of the DEP force, the temperature, in turn, is
simply a measure of the random thermal forces and friction affecting
the colloid. Note that since there were no attractive forces between
the solvent particles and, on the other hand, the solvent density was
much lower than required for hard-sphere crystallization of entropic
origin, the solvent remained in a fluid state even at the lowest
simulation temperatures.

After thermalization the thermostat coupling was retained only within
a distance of 1$\sigma$ from the cell borders for the rest of the
simulation. The border scaling was maintained in order to dissipate
the heat produced from the work done by the electric field on the
colloid, without affecting its DEP motion. A somewhat similar
thermostat coupling scheme was also used by Tanaka and Grosberg in
their simulations of nanoparticle electrophoresis \cite{Tanaka02}. In
those studies it was reported that the thermostat at the cell borders
serves only a minor purpose, as the amount of Joule heat produced
during the simulations was very small. Our test simulations without
the cell border thermostat also showed only negligible increases in
the temperature. We nevertheless retained the border thermostat in the
DEP simulations to dissipate any excess heat in the system.

The equations of motion were integrated using the velocity Verlet
algorithm \cite{Allen_Tildesley}. For proper energy conservation,
integration time steps between 0.002$\tau$ and 0.004$\tau$, depending
on the system temperature, were used.  The total simulation length,
including the system thermalization time, was 5785$\tau$.  During the
simulations the colloid, driven by the DEP force, could move distances
corresponding up to a few simulation cell side lengths (see below in
Sec. \ref{finitesize}). In order to avoid artifacts due to the direct
calculation of the electrostatic interactions and the colloid entering
the thermal scaling region at the simulation cell borders, after each
time step the origin of the simulation cell was set at the interaction
center of the macroion.  In practice, this means that the macroion
itself was not displaced in the simulation cell, but the positions of
all the microions and solvent particles were shifted instead. The
position of the center of the simulation cell in the electric field
coordinates, \vec{R}, was also shifted according to the macroion
motion, as the electric field strength and its gradient changed with
the colloid displacements.

For each combination of electric field strength and temperature, a
total of 20 -- 80 simulations were carried out in order to obtain good
statistics. Larger numbers of simulations were required at higher
temperatures due to thermal noise.

Although the simulations were carried out in the regime of strong
Coulomb coupling (cf. Sec. \ref{elstat}), it is still valid to ask
whether the microions could be stripped from the macroion surface by
random collisions with the solvent particles. In that case the colloid
mobility would be enhanced due to an electrophoretic component, as the
particle would possess a non-zero net charge.  We determined the
minimum microion binding energy for the case of the highest field
strength, $E_0$ = 2.70 \eunit, while the colloidal particle was held
fixed at its initial position in the electric field.  This analysis
showed that even at the highest temperature used in the simulations,
$T$ = 2.4 \tunit, the ratio of the minimum microion electrostatic
binding energy to the thermal energy $k_{\mathrm B} T$ was roughly
24. Hence, the macroion-microion system could be considered stable in
view of collisions with the solvent particles.  All the DEP
simulations were further checked for possible microion detachments,
but no such events were found.

If the electric field-induced colloid motion in the simulations was
truly due to DEP, similar deterministic motion should be obtained
regardless of the actual polarity of the electric field. However, if
the direction of the colloid motion would be reversed by reversing the
field polarity, the motion would be in fact due to electrophoresis. To
make the final validation that the model presented here really
produces DEP motion of the colloidal particle, 10 independent test
simulations were carried out with both positive and negative field
polarities and $\vert E_0 \vert$ = 0.96 \eunit.  In order to
facilitate the comparison, the solvent thermal noise affecting the
colloid was minimized by selecting a very low temperature, $T$ = 0.03
\tunit.  The simulations were otherwise identical in the initial
conditions, except that for the case of negative field polarity, the
microion configurations on the macroion were rotated half a turn with
respect to the radial direction.  Thus, for both field polarities the
colloid was close to their minimum energy state, and delayed coupling
of the colloid to the electric field due to microion distribution
relaxations did not hinder comparison between the two series of
simulations.

In all the cases, regardless of the field polarity, the colloid moved
in the direction of $\nabla \vec{E}^2$, as was expected for positive
DEP.  The average colloid radial displacements for the positive and
negative field polarities were $(-14.0 \pm 2.5)\sigma$ and $(-12.5 \pm
2.5)\sigma$, respectively. Considering the number of test simulations
carried out, it can be stated that the agreement is very good.

\subsection{Finite-size effects}
\label{finitesize}

Before proceeding to the actual DEP simulations, the extent of
finite-size effects in our model should be discussed.  Since in
general hydrodynamic effects, mediated by the solvent, are long-ranged
\cite{Yeh04,Dunweg93}, regardless of the simulation cell size used
there always remains some contribution from interactions between the
colloid and its periodic images. This imposes restrictions on the
minimum size of the solvent shell used in the modeling that in
principle need to be resolved for each specific system.

Probably the best point of comparison in the literature for our model
system is the one used by Tanaka and Grosberg in their MD simulations
of nanoparticle electrophoresis \cite{Tanaka02}. The authors used a
similar WCA solvent and also the same solvent number density as in our
model. The colloid radius in their simulations varied between
3$\sigma$ and 7$\sigma$. From tests on the simulation cell size
dependence the authors concluded that a value of $L_0$ = 20$\sigma$
was sufficient to yield statistically reliable electrophoretic drift
velocities. The actual electrophoresis simulations were then carried
out with $L_0$ = 32$\sigma$.  An important difference between our
simulation system and the one used by Tanaka and Grosberg is that the
latter included co- and counterions in the solvent, which the authors
stated to shield hydrodynamic interactions.  It should be also noted
that in principle the thermostat coupling at the simulation cell
borders we have used could affect the colloid hydrodynamic
interactions. However, as the thermostat coupling was quite weak, this
effect should be very small or even negligible. This is corroborated
by the tests of Tanaka and Grosberg on the cell border thermostat and
nanoparticle electrophoretic drift velocities \cite{Tanaka02}.

\begin{figure}
\centering
\resizebox{0.8\columnwidth}{!}{%
  \includegraphics[angle=-90]{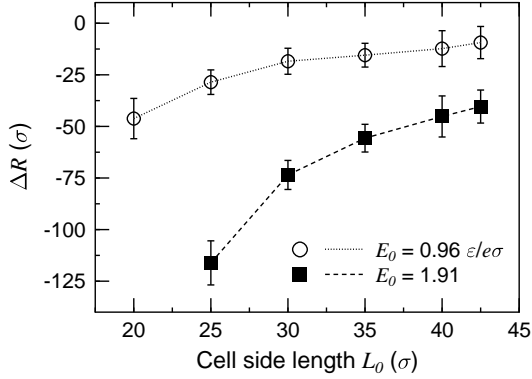}
}
\caption{
Colloid average radial displacements as a function of the simulation
cell side length $L_0$. The error bars in the figure correspond to the
standard deviations of the radial displacements. 
}
\label{sizetest1}       
\end{figure}

We tested several simulation cell sizes in order to find a value of
$L_0$ small enough for good computational efficiency, but still large
enough to result in reasonably small finite-size effects.  As the
measure for an adequate value of $L_0$ we used the colloid average
radial displacements, \averad{}, in reference to the electric field
coordinates, at $E_0$ = 0.96 and 1.91 \eunit. Like in the tests of
opposite electric field polarities (Sec. \ref{simulations}), we used a
temperature of 0.03 \tunit{} to minimize statistical
fluctuations. Figure \ref{sizetest1} shows the results from several
tests employing simulation cell side lengths between $20\sigma$ and
$42.5 \sigma$ with 2239 and 22728 solvent particles, respectively.

For both $E_0$ used the absolute values of the colloid average radial
displacements \averad{} increased with decreasing $L_0$.  This is in
contrast with the well-known enhancement of hydrodynamic friction,
proportional to $L_0^{-1}$ \cite{Yeh04,Dunweg93}. However, standard
hydrodynamic considerations assume an incompressible fluid, which is
not the case with our system.  In the simulations of nanoparticle
electrophoresis, Tanaka and Grosberg also observed \cite{Tanaka02}
increased particle drift velocities for smaller values of
$L_0$. Although in their case the solvent size dependence of the
particle velocity was not smooth as observed here, the effect could
be due to the general properties of the low-density WCA solvent.

A measure of the relative efficiencies of momentum and mass transport
in a fluid at equilibrium is given by the Schmidt number,
\begin{equation} 
Sc = \frac{\eta}{\rho_m D_s},
\end{equation} 
where $\eta$, $\rho_m$ and $D_s$ are the fluid shear viscosity, mass
density, and self-diffusion coefficient, respectively. In real liquids
$Sc$ is typically of the order of $10^3$ and thus momentum propagates
much faster than mass in the medium.  However, for our model solvent,
simple estimates based on the theory of hard sphere fluids
\cite{Hansen_McDonald} result in values of $Sc \approx$ 3 -- 5
(depending on the temperature). This suggests that the apparent
friction on the colloid under non-equilibrium conditions is strongly
connected to the number of solvent particles and hence,
$L_0$. Considering the fact that momentum is not conserved in our DEP
simulations (due to the work done by the electric field on the
colloidal particle and also, to some extent, due to the temperature
scaling only at the simulation cell borders), the exact coupling
between the solvent flow field and the colloid dynamics becomes hard
to assess with analytical means.

Based on the tests, we selected a value of $L_0 = 35\sigma$, with
12687 solvent particles, for the actual DEP simulations.  This value
of $L_0$ still seems to produce larger radial displacements than would
be obtained for a system with a much higher number of solvent
particles.  However, since all the simulations were carried out with
the same system size, and our purpose is to provide qualitative rather
than quantitative insight on DEP, we do not expect finite-size effects
to affect the conclusions of the present work.


\section{Results}
\label{results}

In this Section we present the analysis of the DEP simulations.  In all
the cases the initial position of the colloidal particle was set at a
distance $R_0$ = 1500$\sigma$ on the \xe-axis. Hence, at the start of
the simulations the DEP force drove the colloid in the negative
\xe-direction, while motion in the \ye\ze-plane was perpendicular to
the DEP force.  The DEP force affecting the colloid was varied using
four different electric field strengths with $E_0$ = 0.96, 1.35, 1.91,
and 2.70 \eunit.  Since in our electric field geometry \fdepsc
$\propto E_0^2$, the ratios of the initial DEP forces were
approximately 1:2:4:8. The simulations were carried out at
temperatures between 0.03 and 2.4 \tunit.  The margins of error for
the quantities presented below are estimated from the error of the
mean.

In view of DEP manipulation in the nanoscale, it is instructive to
compare the DEP force to the thermal noise of the surrounding
solvent. For the latter, a characteristic force is given by the ratio
of the thermal energy to the linear size of the particle in question,
\begin{equation}
F_{\rm thermal} = \frac{k_{\rm B} T}{2 a_0},
\end{equation}
where in our case $a_0 \approx r_0 + \sigma/2$ is the radius of the
spherical macroion. A dimensionless parameter describing the ratio of
the thermal noise to the DEP force, is then defined as
\begin{equation}
\Theta_{\rm DEP} = \frac{F_{\rm thermal}}{F_{\rm DEP}} = \frac{
  k_{\rm B} T}{\ap a_0 \arrowvert \nabla E^2 \arrowvert}.
\label{thetadep}
\end{equation}
At small values of \tdep{} the DEP force dominates over the thermal
motion of the colloid, whereas at large values the colloid motion
becomes increasingly random. In the range of electric field strengths
and temperatures used in this study the values of \tdep{} varied
nearly three orders of magnitude, from 0.07 to 40. However, there is
an important point to note here. As the DEP force is strongly
dependent on the radial position in the electric field (\fdepsc
$\propto R^{-5}$), the value of \tdep{} is not fixed. For example, a
colloid that has traversed a distance of 100$\sigma$ radially inward
in the electric field experiences a DEP force that is 41\% larger than
the force at its initial position, $R_0$ = 1500$\sigma$. This further
entails that at sufficiently long times particle distributions in the
radial direction become asymmetric with a negative value of skewness.

It can be questioned whether the DEP force originating from the
polarization of the macroion-microion complex is the same at all
temperatures. As the solvent particles collided with the microions,
the dipole moment of the polarized colloid fluctuated around the value
corresponding to the minimum energy state. However, even at the
highest temperatures used, these deviations were not strong enough to
divert the mean orientation of the dipole moment from the one induced
by the electric field. For all cases the characteristic fluctuation
time of the dipole moment was very short in comparison with the length
of the simulation. It was then plausible to take the average force
affecting the colloid as the one given by Eq. (\ref{sphericalforce}).
Test simulations with a spatially dependent, but non-fluctuating,
artificial force affecting the colloid showed similar radial
displacements as in the DEP case (data not shown), further
corroborating the approximation.

\begin{figure}
\centering
\resizebox{0.85\columnwidth}{!}{%
  \includegraphics[angle=-90]{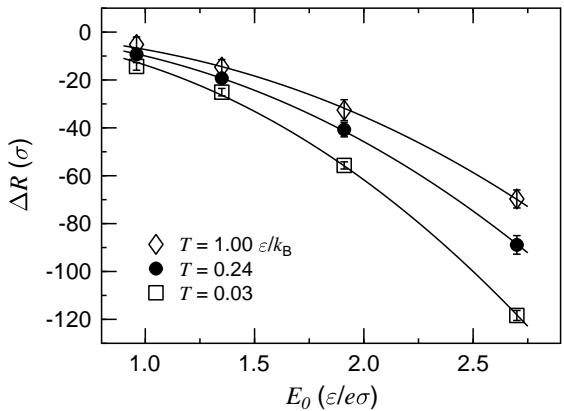}
}
\caption{ 
Average colloid radial displacements, as a function of $E_0$, at the
end of the simulations for three different temperatures.  The solid
lines are the results of the regression analysis between the radial
displacements and the spatially variant DEP force (see text).
}
\label{rdisp_fixt}
\end{figure}

\subsection{DEP at constant temperature}
\label{rdispfixt}

The main results of our simulations were the colloid radial
displacements in the electric field, {\it i.e.}, motion in the
direction parallel to the DEP force.  We first consider series of
simulations where the system temperature was kept constant and $E_0$
was varied. The average radial displacements of the colloid, \averad,
for three different temperatures are shown in Fig.
\ref{rdisp_fixt}. The increase of the radial displacements with the
increase of $E_0$ and the decrease of $T$ is evident.

The ratios of the DEP drift velocities, \avevel{} = \averad/$\Delta t$,
and the colloid thermal velocities at the respective temperatures were
of the order of 0.01 -- 1. At these low values the DEP drift
velocities should be linearly proportional to the DEP force. The
average radial displacement over a time interval $\Delta t$ is then
\begin{equation}
\langle \Delta \vec{R} \rangle = \frac{\vec{F_{\rm DEP}}}{\xi} \Delta t,
\label{fricbalance}
\end{equation}  
where $\xi$ is the friction factor of the solvent. Regression analysis
between the particle radial displacements and the DEP force, $\averad
\propto [\fdepsc]^{\nu}$, justified this assumption. However, for an
accurate analysis it was required to taken into account the fact that
the particle in our simulation is under spatially variant force field:
As the radial position of the colloid changes, so does the magnitude
of the electric field and its gradient. Hence, the radial
displacements were compared to the average DEP force experienced by
the colloid, as calculated from the particle trajectories.  The
resulting values of $\nu$ were, within statistical errors, equal to
unity (see Fig. \ref{rdisp_fixt}).

On the other hand, simply taking the value of the DEP force at the
start of the simulation, \fdepsc($R_0$), resulted in values of $\nu$
slightly higher than one.

\begin{figure}
\centering
\resizebox{0.85\columnwidth}{!}{%
  \includegraphics[angle=-90]{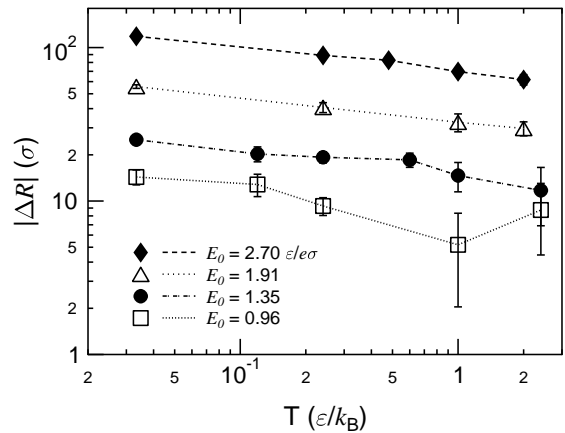}
}
\caption{ 
Absolute values of the average colloid radial displacements for the different
values of $E_0$ as a function of temperature. 
}
\label{rdisp_fixe}
\end{figure}

\subsection{Temperature dependence and DEP mobility}
\label{mobility}

The data in Fig. \ref{rdisp_fixt} indicate that the dependence of the
DEP displacements on temperature is quite weak. We now analyze this
observation in more detail.

Figure \ref{rdisp_fixe} shows the colloid radial displacements, as a
function of temperature, for four series of simulations corresponding
to different values of $E_0$. Note that the data presented are the
absolute values of the radial displacements (in order to make the full
logarithmic plot) but all the real values of the displacements were in
fact negative, {\it i.e.}, in the direction of the DEP force. The data
show a clear power law behavior, although for the two lower values of
$E_0$, at the highest temperatures, the diffusive motion of the
colloid contributes significantly to the scatter in the DEP
displacements. This is characterized by high values of the parameter
$\Theta_{\rm DEP} \approx 8 - 40$.

As the linear response was shown to hold above, a straightforward
assumption is that the friction factor $\xi$ should be linearly
proportional to the solvent shear viscosity $\eta$. An estimate for
the temperature dependence of our solvent viscosity can be obtained
from the well-known Enskog expression for hard-sphere fluids
\cite{Hansen_McDonald}. It has been shown to give good agreement
with MD calculations employing low-density WCA potential solvents
\cite{Bishop83} (such as the one in our case) and further even
qualitatively describe much more complex systems where one would not
necessarily expect the hard-sphere approximation to hold
\cite{Scopigno05}. For the Enskog viscosity the temperature dependence
is $\eta \propto \sqrt{T}$.  Although one could assume even a
qualitative agreement with our low-density WCA solvent friction and
the one given by the Enskog theory, this is clearly not the case
(cf. Fig. \ref{rdisp_fixe}).  What is then the reason for the scaling
of the DEP displacements shown in Fig. \ref{rdisp_fixe}?

Analogous to the case of electrophoresis under uniform \el, it is
possible to define DEP mobility \cite{Morgan99}, \mobdep, that yields
the particle drift velocity due to the electric field,
\begin{equation}
\langle \vec{v}_{\rm DEP} \rangle = \mobdep \nabla E^2.
\label{depmobility}
\end{equation}
Combining Eqs. (\ref{depforce}), (\ref{fricbalance}), and
(\ref{depmobility}) yields the general form
\begin{equation}
\mobdep = \frac{\ap}{2 \xi}.
\label{generalmobility}
\end{equation}
However, care should be taken here. As mentioned above, contrary to
the case where true steady-state condition can be attained, the value
of $\nabla E^2$ changes constantly with particle motion. Calculating
\mobdep{} from large particle displacements and simply using the
initial value of $\nabla E^2$ can lead to significant errors.

\begin{figure}
\centering
\resizebox{0.85\columnwidth}{!}{%
  \includegraphics[angle=-90]{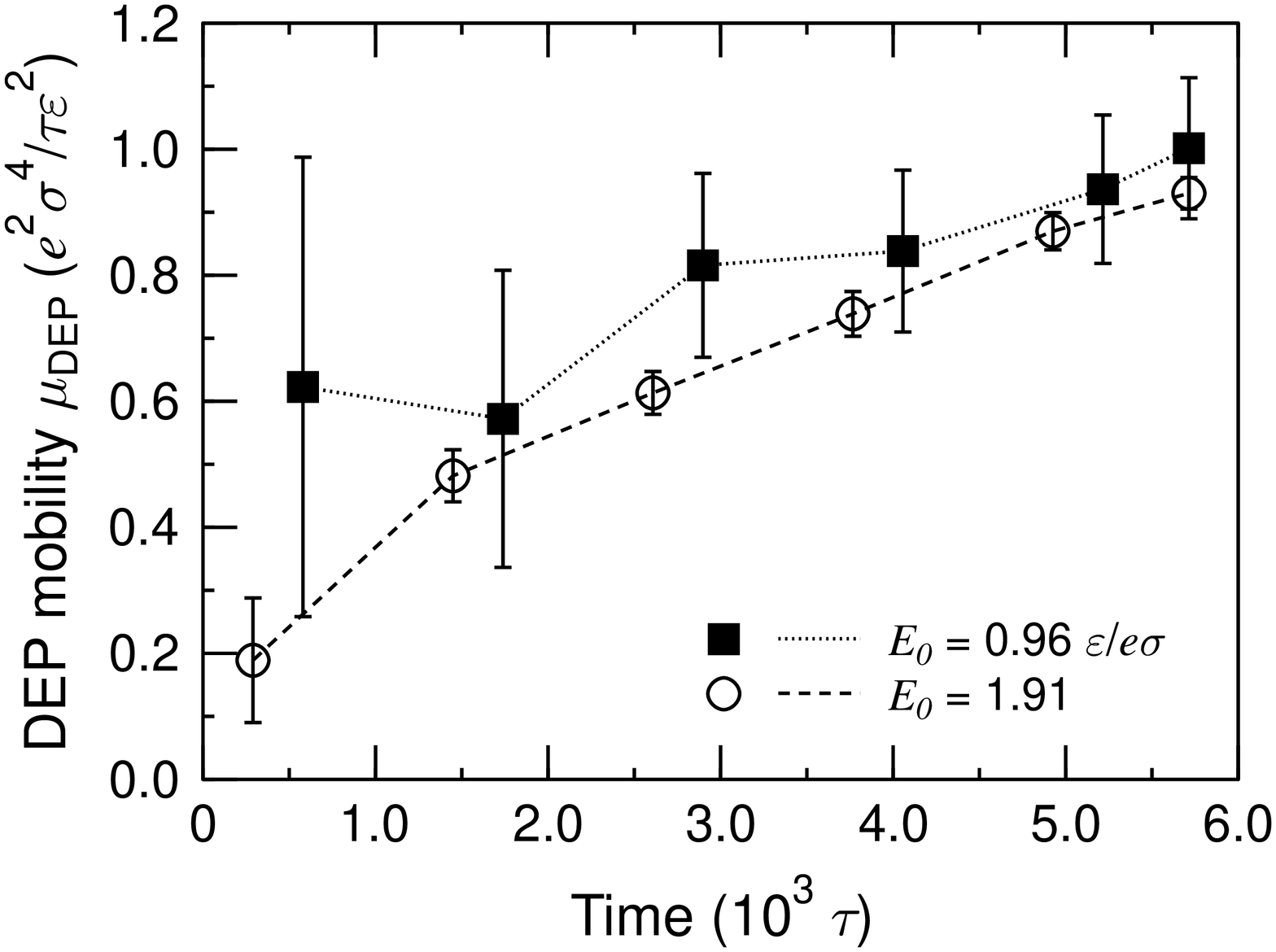}
}
\caption{ 
Colloid mobility \mobdep{} as a function of time for $E_0$ = 0.96 and
1.91 \eunit{} at $T$ = 0.03 $\varepsilon/k_{\rm B}$.
}
\label{mobility10}
\end{figure}

\begin{figure}
\centering
\resizebox{0.85\columnwidth}{!}{%
  \includegraphics[angle=-90]{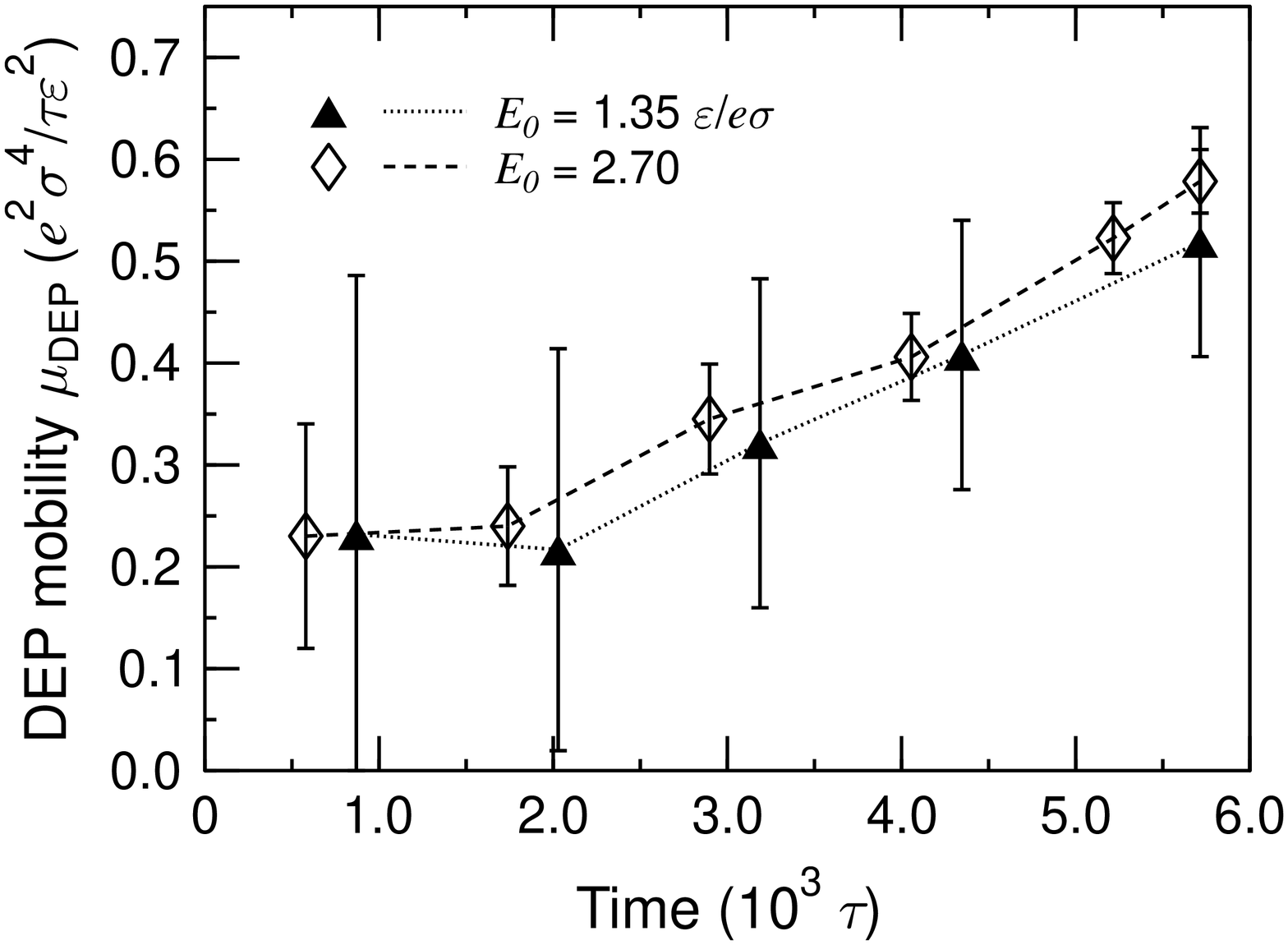}
}
\caption{ 
Same as Fig. \ref{mobility10} but for $E_0$ = 1.35 and 2.70 \eunit{} at 
$T$ = 1.0 $\varepsilon/k_{\rm B}$.
}
\label{mobility300}
\end{figure}

Using Eq. (\ref{depmobility}) we calculated the colloid mobilities
from the particle trajectories and the average values of $\nabla E^2$,
over a given time interval $\Delta t$ from the beginning of the
simulations. The results of these calculations for two different
temperatures are shown in Figs. \ref{mobility10} and
\ref{mobility300}. For clarity, we only show the results for two
different values of $E_0$ in each plot. The data for all $E_0$, at
each respective temperature, were consistent with each other. Also
note that for lower values of $E_0$ the data are somewhat noisier due
to the stronger diffusive motion.

The main observation here is that the values of \mobdep{}, and hence
the values of the friction factor $\xi$ (cf. Eq. (\ref{depmobility})),
{\it vary with time}. Thus, there is a clear coupling of the
non-equilibrium motion of the colloid to the solvent velocity
field. Although a thorough analysis of this effect is challenging, it
is not unreasonable to assume that the characteristic time scale for
the evolution of the non-equilibrium transport (with a constantly
changing friction factor) would be different at different
temperatures.  Indeed, a further study of the {\it apparent}
temperature dependence of the DEP displacements (such as shown in
Fig. \ref{rdisp_fixe} for the final radial displacements in the
simulations) revealed that while power law expressions could be fitted
to the displacement data at different points in time, the resulting
exponents were not constant. We do not want to put too much emphasis
on the actual values of these data, as we were unable to find a
satisfactory theoretical way of interpreting them.

\subsection{Lateral displacements}
\label{lateraldisp}

Another interesting aspect of the spatial dependence of the colloid
motion is seen by studying it in the plane perpendicular to the
electric field gradient at the initial particle position, {\it i.e.},
in the $\ye \ze$-plane. At the beginning of the simulation the
colloidal particle is directed toward the negative \xe-direction by
the DEP force. Deviations from this direction then affect the
precision of the DEP manipulation, {\it i.e.}, how accurately the
final position of the particles can be predetermined.

To a very good approximation the average displacements of the colloid
in the \ye- and \ze-directions were zero for all the combinations of
$E_0$ and $T$.  We then calculated the displacements in the \ye
\ze-plane, \radlat. For comparison, we also determined the colloid
tracer diffusion coefficient $D = (3.65 \pm 0.07) \times 10^{-2}$
$\sigma \sqrt{\varepsilon/m}$ at $T$ = 1.0 \tunit{} in the absence of
the external electric field. This was done using the Green-Kubo
relation between the diffusion coefficient and the particle velocity
autocorrelation function \cite{Hansen_McDonald},
\begin{equation}
D = \frac{1}{3} \int_0^{\infty} \langle \vec{v}(0) \cdot \vec{v}(t)
\rangle \mathrm{d}t.
\end{equation}
A total of 10 independent simulation runs were carried out with a
length of 15800$\tau$ each. The integrals of the velocity
autocorrelation functions were calculated over time periods of
158$\tau$ with a sampling interval of 0.002$\tau$ to ensure proper
convergence.

\begin{figure}
\centering \resizebox{0.8\columnwidth}{!}{%
\includegraphics[angle=-90]{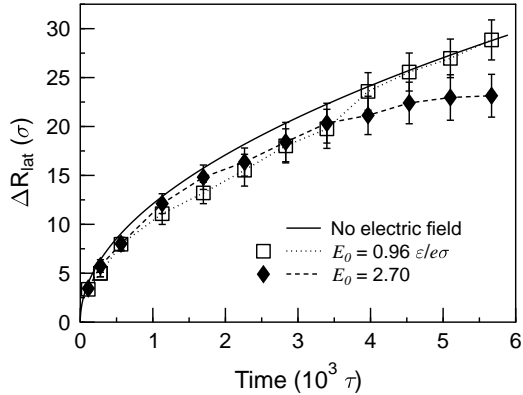} }
\caption{
Lateral (\ye\ze) displacements \radlat{} as a function of time at
$T$ = 1.0 \tunit. The solid line is the prediction obtained from \radlat(t) 
= $\sqrt{4 D t}$.
}
\label{diffusion300}       
\end{figure}

A comparison between the average lateral displacements for the two
extreme cases $E_0$ = 0.96 and 2.70 \eunit, and the root-mean-square
displacement predicted by the relation \radlat = $\sqrt{4 D t}$ is
shown in Fig. \ref{diffusion300}.  In the case of $E_0$ = 0.96 \eunit,
for which \averad{} was small, the colloid average lateral
displacement is in a good agreement with the zero-field diffusion
prediction. However, at $E_0$ = 2.70 \eunit{} a clear difference is
seen. This indicates that with decreasing radial position in the
electric field, the high electric field gradient produces a steering
effect on the colloid. Deviations from motion along the \xe-axis
result in DEP force components also in the \ye- and \ze-directions,
and the assumption of independent motion in the \xe-direction and, on
the other hand, in the \ye\ze-plane breaks down. It is also possible
that the coupling of the particle motion to the solvent velocity field
suppresses the lateral motion up to some degree.

A general relation between the solvent friction and diffusion
coefficient of a particle is given by the Einstein relation
\begin{equation}
D = \frac{k_B T}{\xi},
\label{einstein}
\end{equation}
regardless of the actual mechanism causing the friction. From the
simulated value of $D$ we calculated the friction factor $\xi$ = 27.4
$\pm$ 0.5 $\sqrt{\varepsilon m}/\sigma$, corresponding to the case of
zero external electric field. It is interesting to compare this value
to the time-dependent $\xi$ in the DEP simulations. Inserting the
value of $\xi$ above into Eq. (\ref{generalmobility}) results in
\mobdep{} = (0.090 $\pm$ 0.003) $e^2\sigma^4/\tau\varepsilon^2$ in
equilibrium. This is comparable to the values of \mobdep{} in the
non-equilibrium simulations only at very short times, see
Fig. \ref{mobility300}.  Although not conclusive, the analysis here
implies that at the beginning of the simulation the DEP mobility is
very close to the value obtained by using an equilibrium friction
factor. The coupling to the solvent velocity field then gradually
lowers the solvent friction factor in time and the particle motion
loses its equilibrium nature.


\section{Concluding Remarks}
\label{conclusion}

We have carried out extensive MD simulations of a spherical
nanoparticle DEP in a non-conducting solvent. This study is, to our
knowledge, the first MD study of DEP and serves as a benchmark for
future work employing more complex colloid, solvent, and electric field
models.

The main results of our simulations were the radial displacements of
the colloidal particle due to the DEP coupling. The analysis of these
displacements first showed that the assumption of linear response was
justified for simulation series carried out at constant
temperatures. That is, the DEP drift velocities were linearly
proportional to the DEP force. For an accurate analysis it was
required to take into account the spatial dependence of the electric
field, even at the short time scale of the present simulations.

The comparison of simulation series at different temperatures showed a
weak apparent temperature dependence. We emphasize here again that the
magnitude of this dependence was observed to vary with the time scale
over which the particle motion was studied. 
Furthermore, the particle DEP mobilities, defined by
Eq. (\ref{depmobility}), were observed to increase in time. This is
clearly a manifestation of the non-equilibrium nature of the particle
motion.  The intricate coupling of the particle drifting to the
solvent velocity field decreased the friction factor $\xi$ (which is
inversely proportional to the mobility). A rigorous analytical
treatment of this effect is indeed challenging, and is left as a
subject for future work.

Finally, the analysis of the colloid lateral displacements showed that
while the DEP mobility itself was shown to be time-dependent, for the
lowest value of $E_0$ the colloid motion perpendicular to the initial
direction of the DEP force could still be well predicted by an
equilibrium diffusion coefficient. With increased DEP coupling, this
no longer applied and the lateral motion became coupled to the DEP
drifting.  Intuitively, this is not surprising.  However, it is an
important effect in determining the precision of particle transport,
and we are not aware of any direct quantification of such steering
effects neither in experiments nor computational studies of DEP.

Although thermal motion of the nanoparticles sets limitations on the
DEP forces required for their deterministic motion, other hampering
effects such as the fluid streaming induced by local heating
\cite{Ramos98} can be suppressed with proper experimental set-ups.
For these conditions simple dynamic particle simulations, such as the
ones presented here, can yield valuable insight on the conditions,
precision, and time scales of efficient transport and stable trapping.
This is especially important with particle sizes decreasing to the
nanometer range, where the direct observation of the manipulated
particles and their aggregates becomes very difficult in experiments.


\begin{acknowledgement}
We would like to thank K. W. Yu for many interesting discussions.
This project was supported by the Academy of Finland under project
Nos. 209297 (E.S.), 202598 (E.T.), 80246 (I.V.), and 0019
(M.K.). Grants of computer time from the Finnish IT center CSC in
Espoo, Finland, and the Danish Center for Scientific Computing DCSC at
the University of Southern Denmark are gratefully acknowledged.
\end{acknowledgement}



\end{document}